\documentclass[english,aps,prb,showpacs,twocolumn]{revtex4}
\usepackage[T1]{fontenc}
\usepackage[latin9]{inputenc}
\usepackage{amsbsy}
\usepackage{graphicx}

\providecommand{\tabularnewline}{\\}

\usepackage{babel}

\begin{document}

\title{Quasiparticle and Optical Properties of Rutile and Anatase TiO$_{2}$}

\author{Wei Kang and Mark S. Hybertsen}

\affiliation{Center for Functional Nanomaterials, Brookhaven National Laboratory,
Upton, NY 11973 }
\begin{abstract}
Quasiparticle excitation energies and optical properties of TiO$_{2}$ in the rutile and anatase structures
are calculated using many-body perturbation theory methods.
Calculations are performed for a frozen crystal lattice; electron-phonon coupling is not explicitly considered.
In the GW method, several approximations are compared and it is found that inclusion of the full frequency
dependence as well as explicit treatment of the Ti semicore states are
essential for accurate calculation of the quasiparticle energy band gap.
The calculated quasiparticle energies are in good
agreement with available photoemission and inverse photoemission experiments. 
The results of the GW calculations, 
together with the calculated static screened Coulomb interaction, 
are utilized in the Bethe-Salpeter equation to calculate
the dielectric function $\epsilon_{2}(\omega)$ for both the rutile and anatase structures.
The results are in good agreement with experimental observations, 
particularly the onset of the main absorption features around 4 eV. 
For comparison to low temperature optical absorption measurements that
resolve individual excitonic transitions in rutile, the low-lying discrete excitonic energy levels 
are calculated with electronic screening only.  The lowest energy exciton found in the energy
gap of rutile has a binding energy of 0.13 eV.  In agreement with experiment,
 it is not dipole allowed, but the calculated exciton energy 
exceeds that measured in absorption experiments
by about 0.22 eV and the scale of the exciton binding energy is also too large. 
The quasiparticle energy alignment of rutile is 
calculated for non-polar (110) surfaces. 
In the GW approximation, the valence band maximum is 7.8 eV 
below the vacuum level, showing a small shift
from density functional theory results.
\end{abstract}
\pacs{71.20.Nr 71.15.Qe 71.35.Cc} 
\date{\today}
\maketitle

\section{introduction}

Even after half a century of research\cite{GRANT1959,Diebold2003},
investigation of the fundamental properties of titanium dioxide (TiO$_{2}$) crystal phases
remains important and fruitful, in part due to the
role they have in concepts to effectively utilize solar energy. 
For example, TiO$_{2}$ structures form the photoactive component in
heterogeneous photo-catalysts which, by absorbing energy
from the sunlight, degrade environmentally hazard materials \cite{Fox1993,Hoffmann1995}
and split water into H$_{2}$ and O$_{2}$ \cite{FUJISHIMA1972}.
Scintered anatase TiO$_{2}$ nanoparticles provide the backbone for electron transport and 
the substrate for organic chromophores in
the Gr{\"a}tzel photovoltaic solar cells \cite{OREGAN1991}. In addition to that,
TiO$_{2}$ has been widely used in various areas from optical
coatings to pigments \cite{Diebold2003}. 
Fundamental to all of these applications are the relative alignments
of essential energy levels near the valence and conduction band edges
of TiO$_{2}$ crystal phases and the corresponding optical transition energies.
If predictive computational methods are going to have impact
on the understanding and design of heterogeneous photocatalytic systems based on TiO$_{2}$,
we must first establish that these methods can predict 
the basic properties of the crystal phases, providing a coherent
framework for all the experimental facts. 

Rutile and anatase are two common crystal structures 
in which TiO$_{2}$ is found.
In both phases, each Ti
atom in the crystal is surrounded by a slightly distorted octahedron
formed by six oxygen atoms. The distinct phases exhibit a different connection between
the distorted octahedra (TiO$_{6}$). In the rutile phase
each octahedron shares two edges with its neighbors, while in the
anatase phase each octahedron shares four \cite{Pauling1929}.
In the rutile form, the crystal has a simple-tetragonal structure \cite{Abrahams1971}
with a = b = 0.45936 nm and c = 0.29587 nm. The symmetry of the lattice
is described by the space group $P4_{2}/mnm$ with the only internal
parameter $u$ = 0.30479. In the anatase form the crystal structure
is body-centered tetragonal \cite{Horn1972, Fahmi1993} and
belongs to space group $I4_{1}/amd$. The three sides of the conventional
cell are a = b = 0.3784 nm and c = 0.9515 nm respectively. The internal
parameter $u$ is 0.208 \cite{Horn1972, Fahmi1993}.  
The measurements quoted were done at room temperature;
the change in lattice parameters upon reducing the temperature to 15 K is
less than 0.001 nm \cite{burdett1987}. 

Most of the early first-principles calculations of the properties of TiO$_{2}$ 
were based on the local density approximation (LDA) in a Density Functional
Theory (DFT) based approach \cite{Hohenberg1964,Kohn1965}.
The crystal structures and ground-state
properties were accurately reproduced \cite{glassford1992b,Mo1995,Mikami2000,Asahi2000}.
However, as has been more generally observed for semiconductors and insulators,
the energy gaps pertaining to optical properties were found to be too small.
The minimum energy gap in the LDA band structure underestimated
the band gap observed
in optical experiments \cite{Pascual1977,Tang1993}
by about 40\% \cite{Mikami2000}. 
Calculations based on Hartree-Fock theory have been performed,
giving accurate structural properties for rutile and anatase, but with 
a minimum energy gap that exceeded 10 eV \cite{Fahmi1993}. 
A hybrid approach, admixing a fraction of the bare exchange from Hartree-Fock,
also showed accurate structural properties for rutile \cite{Zhang2005},
with a band gap that is closer to experiment (3.4 eV) \cite{Muscat2001}.
A more economical approach, approximately accounting for explicit
Coulomb interactions through a U parameter acting on the Ti 3d electrons
in a DFT+U approach,
overestimated the lattice parameters while still showing a band gap that
was smaller than experiment \cite{Nolan2008}.

A direct approach to calculate
electronic excitation energies based on the Green's function approach of many-body
perturbation theory (MBPT), specifically utilizing
the GW approximation for the electron self energy \cite{Hedin1965},
has proven to be relatively accurate for a broad array of materials \cite{Hybertsen1986,Aryasetiawan1998,Aulbur19991,Onida2002}.
Several calculations have been reported for TiO$_{2}$
with different implementations of the GW method \cite{Oshikiri2003,Schilfgaarde2006,Thulin2008,Mowbray2009}. 
However, the band gap was significantly overestimated in all
these reports. 

The frequency dependent macroscopic dielectric function
probed in optical measurements has been extensively studied within the
framework of DFT for TiO$_{2}$ crystal phases \cite{Glassford1992,glassford1992b,Mo1995,Asahi2000}.
Using the independent particle approach of
Ehrenreich and Cohen \cite{Ehrenreich1959}, 
the underestimate of the fundamental band gap immediately gives an error
in the optical threshold and 
the overall shape of the dielectric function
calculated in this way was quite different from the experimental measurements
\cite{Cardona1965,Hosaka1997}.  
In the framework of MBPT, the GW-based results for the electronic
excitation energies are input to a direct treatment of neutral excitations
through solution of the two-particle Bethe-Salpeter
equation (BSE), an approach that has provided
a satisfactory description of the optical properties of a number of systems \cite{Rohlfing2000,Onida2002}.
An application of the BSE approach for rutile and anatase TiO$_{2}$ 
has only recently appeared in the literature \cite{Lawler2008}.
The shape of the spectra are in much better agreement with experiment 
as compared to the independent particle approach.

In this article, we critically assess the application of MBPT methods
to calculate the electronic and 
optical excitations for TiO$_{2}$ in the rutile and anatase crystal phases.
To treat the electronic excitation energies, we use the GW method
without self consistency or inclusion of vertex corrections.
Empirically, this approach is often relatively accurate, 
although a full understanding of cancellations between self consistency
and vertex corrections remains 
an open area of research \cite{Shirley1996,Holm1998,Ku2002,
Bruneval2006,Kotani2007,Shishkin2007a,Shishkin2007b}.
Several different approximations to handle the frequency
dependence of the screened Coulomb interaction in the GW method
are compared.  We find that use of plasmon pole models \cite{Hybertsen1986,Linden1988}
results in significant overestimation of the band gap.
When the full frequency dependence of the screening is included,
together with explicit treatment of the Ti semicore states,
the calculated electronic excitation spectrum is found to
be in good agreement with photoemission and inverse photoemission spectra \cite{Tezuka1994,see1994}.
Interestingly, the calculated fundamental gap (3.34 eV and 3.56 eV for rutile and anatase respectively),
is still larger than the measured minimum gap from optical absorption
(3.03 eV for rutile \cite{Pascual1977} and estimated to be 3.3 eV for anatase \cite{Tang1993}).
As a first step towards application to heterogeneous photocatalytic systems,
we discuss the alignment of the valence and conduction band edges
at non-polar vacuum-solid (110) interface of TiO$_{2}$.
We find that the GW method implies only very small corrections relative to
the LDA for the valence band 
position, with the most of the band gap error going to shift the conduction band edge upwards.

To explore the role of electron-hole interactions and excitonic binding
energy, we have used the results from the GW based calculations
as input to the BSE approach.
Similar to the recent results of Lawler $\it{et}$ $\it{al.}$ \cite{Lawler2008},
the calculated frequency dependent dielectric function accurately
reproduces the main onset of absorption near 4 eV and gives
a good account of the frequency dependence for both rutile and anatase.
We also solve the BSE for the low-lying, bound exciton states for rutile.
The deepest exciton binding energy is calculated to be about 0.13 eV.
The dipole-forbidden character of the lowest exciton agrees
with low temperature measurements \cite{Pascual1977},
although the predicted exciton energy (3.25 eV) is still 0.22 eV
larger than experiments.
Also, the magnitude of the exciton binding energy is larger.
While the discrepancy for the exciton energy 
could very reasonably be regarded as within the
expected errors of the MBPT methods used here, it may suggest
an important role for electron-phonon coupling
in screening and in further
renormalizing the energy gap in TiO$_{2}$.

The rest of the article is organized as follows: In Sec. II, the
methodologies used in the DFT, electronic excitation and optical excitation calculations are
briefly summarized, the key approximations are discussed and
the numerical details are provided.
In Sec. III, we present the main results for the electronic
and optical excitations in rutile and anatase TiO$_{2}$ 
and discuss them in comparison to available experiments.
Finally, we conclude the article
in Sec. IV with a short discussion, including the role of coupling to phonons.

\section{Methodology and Numerical Details}

\subsection{DFT Calculations}

The LDA eigenvalues and eigenvectors of TiO$_{2}$ are calculated
with a plane-wave basis set using norm-conserving pseudopotentials.  Unless indicated
otherwise, the LDA calculations are carried out using the ABINIT package\cite{Gonze2002,Gonze2005}.
In TiO$_{2}$ the Ti is nominally ionized to [Ti$^{4+}$] and the
low lying conduction band states are of predominantly $3d$ character.
As we show below, artificially dividing the n=3 shell of Ti into frozen core ($3s$ and $3p$) and
valence ($3d$) contributions introduces a significant error to the energy band gap.
The pseudopotential of Ti which includes semicore electrons is generated
using the OPIUM package \cite{opium} in the Troullier-Martins scheme \cite{Troullier1991}
with an initial configuration of $(Ne)3s^{2}3p^{6}3d^{0}4s^{0}4p^{0}$.
The outermost five orbitals are included and the cutoff radii (in
Bohr) are 0.9, 0.9, 1.0, 0.9, and 0.9 respectively. Other
pseudopotentials are taken from the ABINIT pseudopotential database
generated using the FHI99PP package \cite{Fuchs1999}. 

In all calculations,
the Perdew-Wang representation \cite{Perdew1992} of Ceperly-Alder
exchange-correlation potential \cite{Ceperley1980} is used. 
When including the Ti semicore states, 
a kinetic energy cutoff of 200 Ry is used to ensure the convergence
of the LDA results, as suggested by previous calculations\cite{Mikami2000,Vast2002}.
To examine the accuracy of the pseudopotentials, we calculate the
optimized lattice constants for rutile, 
finding $a~=$ 4.5484 \AA  (4.5936 \AA), $c/a~=$ 0.6414 (0.64409)
and $u~=$ 0.3040 (0.30479), agreeing with the experimental values
noted in parentheses \cite{Abrahams1971} to the accuracy generally expected for LDA calculations.
We also compare our LDA calculations with results obtained using the VASP package \cite{Kresse1996,Kresse1996b}
with the recommended projector augmented-wave (PAW) pseudopotentials \cite{Blochl1994b}.
The difference between the two LDA calculations is within 0.5\% for lattice parameters and less than
0.05 eV for bandgaps.
In the GW and BSE calculations described below, 
the geometrical parameters of the unit cell for both rutile and anatase
phases are taken from
experimental measurements\cite{Abrahams1971,Cromer1955,Fahmi1993,Horn1972}.

\subsection{GW Method}

In MBPT, the evolution of the electrons in a material is described
by the one-particle Green's function, with the effect of electron-electron interactions
represented by an electron self energy operator.
Well defined electronic excitations appear as peaks in the
corresponding spectral function.
Excitations with single particle character, namely quasiparticles, can be
obtained as solutions of a Schrodinger-like
equation \cite{Hedin1969}
\begin{eqnarray}
(T+V_{ext}+V_{H})\psi_{n,k}\left(\boldsymbol{r}\right)+\qquad\nonumber \\
\int d\boldsymbol{r}'\Sigma(\boldsymbol{r},\boldsymbol{r}';E_{n,k})
\psi_{n,k}(\boldsymbol{r}') & = & E_{n,k}\psi_{n,k}
\quad\label{eq:1}
\end{eqnarray}
where $T$ is the kinetic energy, $V_{ext}$ is the external potential,
and $V_{H}$ is the average Hartree potential. 
$\Sigma$ is the self
energy of the electrons and the indices refer to Bloch states n, k. It includes all the exchange-correlation
effects contributed by surrounding electrons. 
Since $\Sigma$ is generally
non-Hermitian, $E_{n,k}$ is complex with the real part giving the quasiparticle energy
and the imaginary part corresponding to the
width of the quasiparticle peak in the spectral function, i.e. the quasiparticle lifetime. 

A practical approximation to calculate $\Sigma$
has proven to be the so-called GW approximation of Hedin \cite{Hedin1965}, in which the self energy $\Sigma(\boldsymbol{r,}\boldsymbol{r}';E)$
is formally written as
\begin{eqnarray}
\Sigma(\boldsymbol{r},\boldsymbol{r}';E)=\qquad\qquad\qquad\qquad\qquad\qquad\qquad\qquad\nonumber \\
\frac{i}{2\pi}\int dE'e^{-i\delta^{+}E'}G(\boldsymbol{r},\boldsymbol{r}';E-E')W(\boldsymbol{r},\boldsymbol{r}';E'),\label{eq:2}
\end{eqnarray}
Here $G$ is the Green's function of the electrons and $W$ is the dynamically
screened Coulomb interaction determined by the inverse dielectric matrix $\epsilon^{-1}(\boldsymbol{r},\boldsymbol{r}'';E)$, and $\delta^{+}$ is a positive infinitesimal time. The $G$ and $W$ in Eq. (\ref{eq:2}) refer
to the fully interacting Green's function.  
However, in practice,
using an initial LDA calculation to determine the screening through linear response calculations (not including the exchange-correlation
kernel) and to provide an initial, independent-particle Green's function has often proven
to be sufficiently accurate.
There are specific examples where the LDA orbital character can be wrong, e.g. in some late $3d$
transition metal compounds\cite{Aryasetiawan1995,Faleev2004,Bruneval2006a}. 
However, in TiO$_{2}$, the Ti $3d$ is almost empty
and the valence band edge region is predominantly O $2p$ character
with minimal admixture of Ti $3d$.  The TiO$_{2}$ case should be similar to the
vast majority of semiconductors and insulators in this regard.  Also, the LDA wavefunctions are
sufficiently accurate
that a first order estimate of the self energy correction to the LDA eigenvalues is adequate.
The quasiparticle energy correction
$\Delta E_{n,k}$ to a LDA orbital $\phi_{n,k}$ is obtained
through a reduced form of Eq. (\ref{eq:1}) as
\begin{equation}
\Delta E_{n,k}=Z_{n,k}\left\langle \phi_{n,k}|\Sigma(E_{n,k}^{LDA})-V_{xc}^{LDA}|\phi_{n,k}\right\rangle ,\label{eq:4}
\end{equation}
 where $V_{xc}^{LDA}$ is the exchange-correlation potential and $Z_{n,k}$
is the renormalization factor of the orbital defined as $Z_{n,k}=(1-\partial\Sigma/\partial E)^{-1}|_{E=E_{n,k}^{LDA}}$. 

The frequency dependence of the screened Coulomb interaction
($W$) can often be addressed 
using a generalized plasmon-pole (GPP) model \cite{Hybertsen1986,Linden1988},
with substantial advantages in computational efficiency.
The GPP models have proven to be relatively accurate
for many semiconductors and insulators, including ionic crystals
such as LiCl \cite{Hybertsen1985} and MgO \cite{Shirley1998}. 
However, as discussed below, we find that use of the GPP
leads to a gap that is substantially too large for TiO$_{2}$.
Several approaches to include the full frequency-dependent
dielectric matrix have been implemented and described
in the literature: (1) an analytical continuation method \cite{Rojas1995,Rieger1999},
(2) a direct method which carries out the integration in Eq. (\ref{eq:2})
along the real axis \cite{Marini2002,marini2002b,Shishkin2006}, and
(3) a contour deformation (CD) method which deforms the integration
in Eq. (\ref{eq:2}) along the imaginary axis \cite{Leb`egue2003}.
We adopt the contour deformation method to carry out the calculations,
which is particularly efficient for evaluating self energy for
states near the gap region. 

In the CD method, the correlation contribution $\Sigma^{c}(\boldsymbol{r},\boldsymbol{r}';E)$
of the self energy is written as the sum of two terms \cite{Leb`egue2003, bruneval2005} 
\begin{widetext}
\begin{eqnarray}
\Sigma^{c}(\boldsymbol{r},\boldsymbol{r}';E)  =  
-\sum_{n,k}\phi_{n,k}(\boldsymbol{r})\phi_{n,k}^{*}(\boldsymbol{r}')
\left\{ \frac{1}{\pi}\int_{0}^{\infty}dE''\frac{E-E_{n,k}}{(E-E_{n,k})^{2}+E''^{2}}W_{p}
(\boldsymbol{r},\boldsymbol{r'},iE'') \right. \quad \quad & & \nonumber \\
  +  \left.W_{p}(\boldsymbol{r},\boldsymbol{r}';|E-E_{n,k}|-i\eta)
\left[\Theta(E_{f}-E_{n,k})\Theta(E_{n,k}-E)-\Theta(E_{n,k}-E_{f})
\Theta(E-E_{n,k})\right]\right\}, & & \nonumber \label{eq:5} \\ 
\end{eqnarray}
\end{widetext}
where $W_{p}=W-V_{coul}$, $E_{f}$ is the Fermi energy, $\eta$ is
a small damping amplitude and $\Theta$ is a Heaviside function. 
The first term in Eq. (\ref{eq:5}) comes from the integration along the
imaginary axis. As W is now smooth along the imaginary axis, a sparse
sampling of $E$ is sufficient to converge the integration. The second
term is the residual contribution of poles near the real axis. It
is non-zero only while $E>E_{n,k}>E_{f}$ or $E<E_{n,k}<E_{f}$.
For any $E$ close to the Fermi surface, only $W_{p}$ for $|E-E_{n,k}|\sim0$
have non-vanished contributions to $\Sigma^{c}$ in the second term. This makes the calculation 
more computational efficient; $W$ is a smooth function of
$|E-E_{n,k}|$ around 0, due to the band gap,
and only relatively low frequencies need to be sampled. 

We implement the contour deformation approach
based on a private branch of the YAMBO package \cite{Marini2009}.
The integration along the imaginary axis in Eq. (\ref{eq:5}) is performed
with a non-uniform mesh of $N$ points according to
\begin{equation}
E_{i}''=E_{0}\tan(\frac{i-1}{2N}\pi),\qquad i=1,2,...,N,\label{eq:9}
\end{equation}
which maps the integration along the imaginary axis to an integration
on the {[}0, 1) interval. The energy $E_{0}$ provides a scale for the overall 
density 
of the samples on the imaginary axis. Half of the mesh spans the energy scale
from zero up to $E_{0}$
while the other half sample the higher energies. For TiO$_{2}$, a mesh
of 50 points and an energy scale of 1 Ry were enough to keep the numerical
error of the integration within 1 meV. $W_{p}$ on the real axis is
uniformly sampled with an energy increment of 0.1 eV and values between
mesh points are linearly interpolated. 
The  special case
in Eq. (\ref{eq:5}) for $E \rightarrow E_{n,k}$ must be handled with care. 
A consistent treatment, that avoids the apparent divergence
and properly handles all the terms in Eq. (\ref{eq:5}), 
is to add a small positive energy to $E - E_{n,k}$
(say $\delta=2.0\times10^{-7}$ Ry) when necessary. 
A very similar contour deformation approach has been implemented
in ABINIT \cite{bruneval2005}, and we have carefully compared the
results for test cases.  The ABINIT package has also been used for the 
GPP model calculations.
 
For all the GW calculations, the energy cutoff
is 60 Ry for the evaluation of the bare Coulomb exchange contribution
$\Sigma_{x}$ , and 20 Ry for the correlation part $\Sigma_{c}$.
A total of 160 bands are used for the calculation of both dielectric matrices
and self-energies. An unshifted $4\times4\times6$ Monkhost-Pack (MP)
mesh\cite{Monkhorst1976} is used to sample the Brillouin zone (BZ)
of rutile, while for anatase an unshifted $4\times4\times4$ MP mesh
is used.  

\begin{figure}
\includegraphics[clip,scale=0.35]{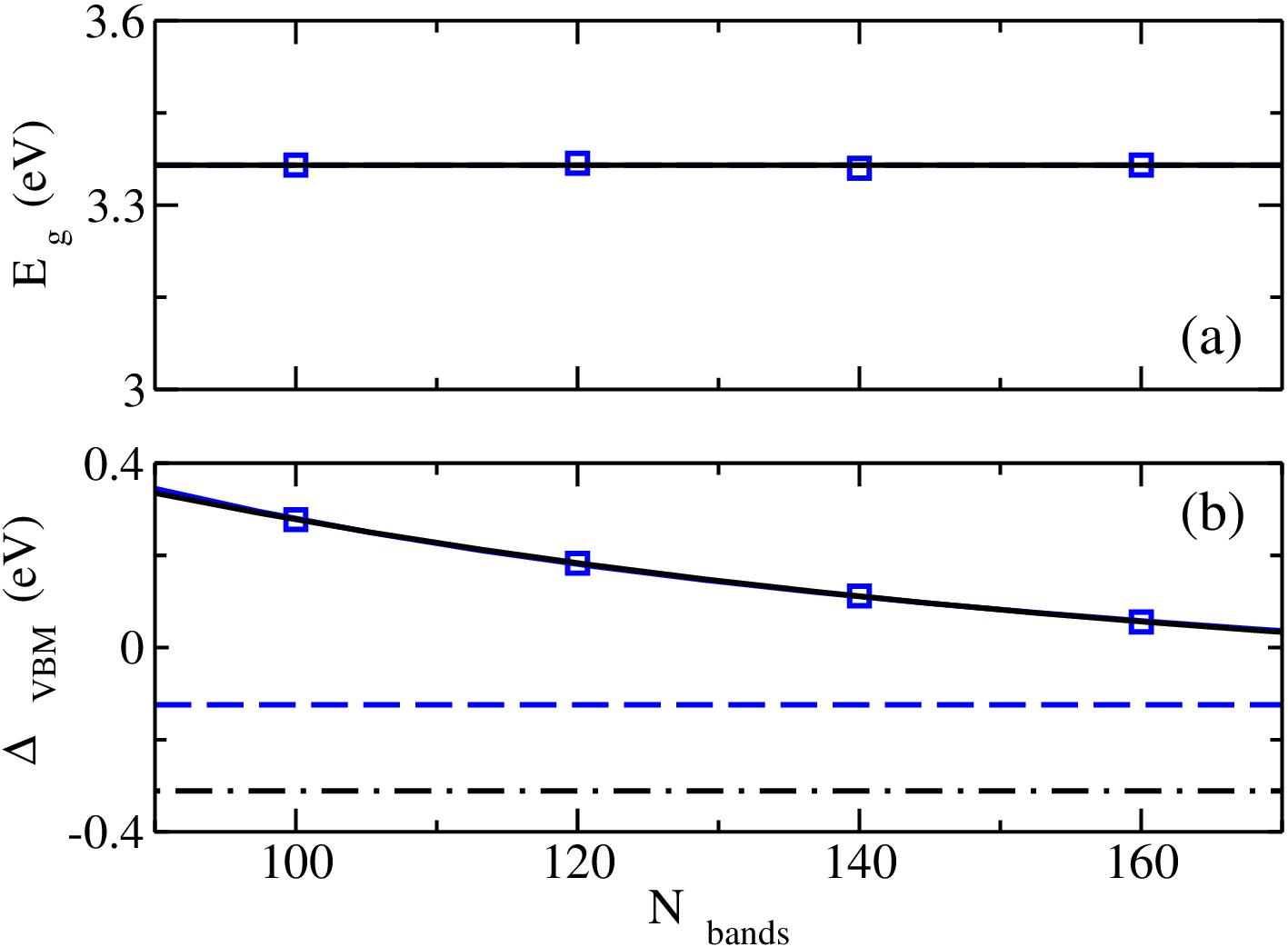}

\caption{Quasiparticle direct energy gap at $\Gamma$ (a) 
and  energy of VBM (b) for rutile as a function
of the total number of bands kept in the full frequency dependent GW calculation.
Squares represent calculations with Ti semi-core electrons included explicitly
as valence electrons. The results are
fitted using two different functional forms described in the text
and displayed as solid lines that are indistinguishable in the figure.
However, in (b), they have different $N \rightarrow \infty$ limits,
displayed as dash-dot ($1/N$) and long dash ($exp(-N/c)$) lines in
the figure respectively.  \label{fig:0}}

\end{figure}

A test of the convergence with respect to the number of bands included
is shown in Fig. \ref{fig:0} for the final full-frequency approach with
Ti semicore electrons treated explicitly as valence electrons in the pseudopotential.
In order to characterize the fully converged values, the data were
fit with two different empirical forms, $E(N)=E_{0}-b/N$, and $E(N)=E_{0}-b\cdot exp(-N/c)$.
We first check the validity of the fitting forms for the case of bulk silicon.
The exponential form closely represents the band edge and band gap
energies as a function of the number
of included conduction bands, yielding extrapolated results
in excellent agreement with those obtained via methods suggested by
Bruneval and Gonze\cite{bruneval2008}. For rutile, the fitting
curves displayed in Fig. \ref{fig:0} are indistinguishable, but predict
slightly different $N\rightarrow\infty$ results for the absolute
shift of the valence band edge, as indicated by the horizontal dashed
lines in Fig. \ref{fig:0}(b). In particular, the fit for the quasiparticle
energy gap indicates a converged quasiparticle energy of 3.37
eV for N=160. For the absolute shift of the valence band edge, the
convergence is somewhat slower, with extrapolated values of  -0.12 eV and -0.31 eV. This suggests that
the valence band edge in the final results is  0.2 to 0.4 eV lower than the N=160 value.

\subsection{Bethe-Salpeter Equation and Optical Properties}

A detailed description of the
BSE method is given in the literature\cite{Onida2002,Rohlfing2000}.
We use the implementation in the public branch of the YAMBO package \cite{Marini2009}.
With the Tamm-Dancoff  approximation \cite{Fetter1971} and the use of a 
static screened Coulomb interaction, the BSE is simplified to a generalized eigenvalue equation
$H_{vc,v'c'}A_{s}^{v'c'}=E_{s}A_{s}^{vc}$. 
The effective Hamiltonian
$H_{vc,v'c'}$ has an explicit form of
\begin{equation}
H_{vc,v'c'}=(E_{c}-E_{v})\delta_{vv'}\delta_{cc'}+2\bar{V}_{vc,v'c'}-\bar{W}_{vc,v'c'},\label{eq:7}
\end{equation}
where the quasiparticle energies (taken from the GW calculations) enter on the diagonal,
$\bar{V}_{vc,v'c'}=\left\langle vc\left|V_{Coul}\right|v'c'\right\rangle $, 
$\bar{W}_{vc,v'c'}=\left\langle vv'\left|W(E=0)\right|cc'\right\rangle $
and the indices $v,~c$ refer to the occupied valence and empty conduction 
band states. For brevity, the explicit reference to Bloch 
wavevector $k$ for each state is suppressed.
The effective Hamiltonian
in Eq. (\ref{eq:7}) is explicitly written for the spin singlet excitations
which are directly probed in optical measurements.
For the spin triplet excitations, the effective Hamiltonian is modified
by dropping the so-called exchange term $2\bar{V}_{vc,v'c'}$.
In terms of $E_{s}$ and $A_{s}^{vc}$, the macroscopic dielectric
function $\epsilon_M(\omega)$, including local field effects \cite{Rohlfing2000,Onida2002}, is expressed as
\begin{equation}
\epsilon_M(\omega)=1-\lim_{q\rightarrow0}\frac{4\pi e^{2}}{q^{2}}\sum_{s}\frac{\left|{\displaystyle \sum_{vc}}\left\langle v\left|e^{-i\boldsymbol{q}\cdot\boldsymbol{r}}\right|c\right\rangle A_{s}^{vc}\right|^{2}}{\omega-E_{s}+i\eta}.\label{eq:8}
\end{equation}
In YAMBO, we use the option to evaluate the response function recursively \cite{Benedict1999,Marini2009}. 
In order to study specific, low energy exciton states, we also 
directly diagonalize the generalized eigenvalue equation.

Calculations of optical properties via BSE are more expensive computationally.
For both phases, the static dielectric matrices are calculated with
80 bands and a damping coefficient of 0.1 eV. 
The dielectric function $\epsilon(\omega)$ is calculated on an $8\times8\times12$
MP mesh for rutile and on an $12\times12\times12$ MP mesh for anatase.
For both cases, electron-hole
pairs within 15 eV are taken in to build up the BSE kernel, in which
the energy cutoff is 10 Ry for $\bar{V}$ and 3.5 Ry for $\bar{W}$.
To calculate the excitonic
binding energy of rutile, we restrict the basis set
for the effective Hamiltonian to one conduction band and one valence band.
The exciton binding energy converges relatively slowly with BZ sampling.
The final results are reported based on a $12\times12\times18$ MP mesh to
sample the BZ.  
The energy cutoff is larger in this calculation for a more accurate representation of the BSE kernel,  14 Ry for $\bar{V}$ and 6
Ry for $\bar{W}$.

\begin{table*}
\begin{tabular}{cccccccccccccccccccccccccccccccccccccccccccccc}
\hline 
\noalign{\vskip\doublerulesep}
 & Method &  &  &  & SC &  &  &  & \multicolumn{10}{c}{VBM} &  &  &  &  &  &  & \multicolumn{10}{c}{CBM} &  &  &  &  &  &  & \multicolumn{4}{c}{E$_{g}^{\Gamma}$} & \tabularnewline[\doublerulesep]
\noalign{\vskip\doublerulesep}
\noalign{\vskip\doublerulesep}
 &  &  &  &  & in PP &  &  &  & V$_{xc}$ &  &  & $\Sigma_{x}$ &  &  & $\Sigma_{c}$ &  &  & Z &  &  &  &  &  &  & V$_{xc}$ &  &  & $\Sigma_{x}$ &  &  & $\Sigma_{c}$ &  &  & Z &  &  &  &  &  &  & LDA &  &  & GW & \tabularnewline[\doublerulesep]
\hline
\noalign{\vskip\doublerulesep}
\noalign{\vskip\doublerulesep}
 &  &  &  &  &  &  &  &  &  &  &  &  &  &  &  &  &  &  &  &  &  &  &  &  &  &  &  &  &  &  &  &  &  &  &  &  &  &  &  &  &  &  &  &  & \tabularnewline[\doublerulesep]
\noalign{\vskip\doublerulesep}
\noalign{\vskip\doublerulesep}
 & GPP(HL) &  &  &  & n &  &  &  & -19.97 &  &  & -24.59 &  &  & 3.94 &  &  & 0.82 &  &  &  &  &  &  & -11.48 &  &  & -3.57 &  &  & -5.69 &  &  & 0.84 &  &  &  &  &  &  & 2.08 &  &  & 4.48 & \tabularnewline[\doublerulesep]
\noalign{\vskip\doublerulesep}
\noalign{\vskip\doublerulesep}
 & GPP(VDLH) &  &  &  & n &  &  &  & -19.97 &  &  & -24.59 &  &  & 4.37 &  &  & 0.82 &  &  &  &  &  &  & -11.48 &  &  & -3.57 &  &  & -6.35 &  &  & 0.85 &  &  &  &  &  &  & 2.08 &  &  & 3.60 & \tabularnewline[\doublerulesep]
\noalign{\vskip\doublerulesep}
\noalign{\vskip\doublerulesep}
 & FF(CD) &  &  &  & n &  &  &  & -19.97 &  &  & -24.59 &  &  & 4.50 &  &  & 0.71 &  &  &  &  &  &  & -11.47 &  &  & -3.57 &  &  & -5.78 &  &  & 0.73 &  &  &  &  &  &  & 2.08 &  &  & 3.73 & \tabularnewline[\doublerulesep]
\noalign{\vskip\doublerulesep}
\noalign{\vskip\doublerulesep}
 &  &  &  &  &  &  &  &  &  &  &  &  &  &  &  &  &  &  &  &  &  &  &  &  &  &  &  &  &  &  &  &  &  &  &  &  &  &  &  &  &  &  &  &  & \tabularnewline[\doublerulesep]
\noalign{\vskip\doublerulesep}
\noalign{\vskip\doublerulesep}
 & GPP(HL) &  &  &  & y &  &  &  & -20.21 &  &  & -24.67 &  &  & 4.09 &  &  & 0.83 &  &  &  &  &  &  & -20.37 &  &  & -12.13 &  &  & -5.64 &  &  & 0.85 &  &  &  &  &  &  & 1.75 &  &  & 4.27 & \tabularnewline[\doublerulesep]
\noalign{\vskip\doublerulesep}
\noalign{\vskip\doublerulesep}
 & GPP(VDLH) &  &  &  & y &  &  &  & -20.21 &  &  & -24.67 &  &  & 4.41 &  &  & 0.83 &  &  &  &  &  &  & -20.37 &  &  & -12.13 &  &  & -5.99 &  &  & 0.85 &  &  &  &  &  &  & 1.75 &  &  & 3.70 & \tabularnewline[\doublerulesep]
\noalign{\vskip\doublerulesep}
\noalign{\vskip\doublerulesep}
 & FF(CD) &  &  &  & y &  &  &  & -20.21 &  &  & -24.67 &  &  & 4.56 &  &  & 0.70 &  &  &  &  &  &  & -20.37 &  &  & -12.12 &  &  & -5.88 &  &  & 0.72 &  &  &  &  &  &  & 1.75 &  &  & 3.38 & \tabularnewline[\doublerulesep]
\noalign{\vskip\doublerulesep}
\noalign{\vskip\doublerulesep}
 &  &  &  &  &  &  &  &  &  &  &  &  &  &  &  &  &  &  &  &  &  &  &  &  &  &  &  &  &  &  &  &  &  &  &  &  &  &  &  &  &  &  &  &  & \tabularnewline[\doublerulesep]
\hline
\noalign{\vskip\doublerulesep}
\end{tabular}

\caption{Analysis of the valence band maximum (VBM), the conduction band minimum (CBM)
and the direct energy gap $E_{g}^{\Gamma}$
at the $\Gamma$ point of the Brillouin zone calculated for rutile TiO$_{2}$
using the GW method with several different approximations.
The methods refer to two different generalized plasmon-pole
(GPP) models and the full frequency-dependent (FF) approach described in the text.
The second column indicates whether Ti semicore states are explicitly included
as valence electrons.
For the VBM and CBM, the expectation value is shown for the 
exchange-correlation potential $V_{xc}$ in the LDA,
the bare exchange self energy ($\Sigma_{x}$), 
the correlation part of the self energy ($\Sigma_{c}$)
and the renormalization factor ($Z$).
The band gap is shown in the LDA and for the GW method for each case.
Energies are given in eV and the renormalization factor $Z$ is dimensionless.\label{tab:1} }

\end{table*}

\section{Results}

\subsection{Electronic Excitation Energies in TiO$_{2}$}

The calculated electronic excitation energies in titanium oxides 
are found to be sensitive to technical factors in the GW calculations.
We illustrate that here for the case of TiO$_{2}$ in the rutile phase (Table \ref{tab:1}).
First, explicit, self consistent treatment of the semicore electrons ($3s$ and $3p$) on the Ti
in the calculations for the solid affects the calculated energy gaps.
As discussed in the literature, although the $3s$ and $3p$ levels are well separated
from the $3d$ states energetically, there is significant spatial overlap \cite{marini2002b}. 
The effect for TiO$_2$
is already evident at the LDA level, where the gap is reduced by more than 0.3 eV
upon explicit inclusion of the semicore electrons
relative to freezing the semicore electrons in the pseudopotential.

Depending on the approximation used to treat the electron self energy
in the GW method, the net influence of the semicore electrons varies.
Focusing on the influence of the semicore electrons for the most accurate, 
full-frequency method (FF), the change is greater than 0.3 eV. 
A more detailed view of the contributions of
the LDA exchange-correlation potential $V_{xc}$, the exchange $\Sigma_{x}$
and the correlation $\Sigma_{c}$ part of the self energy are also displayed
Table \ref{tab:1}. 
The valence band maximum (VBM) orbitals are largely composed of
oxygen $p$-states.
They are spatially separated from the Ti semicore electrons, so the changes are relatively small.
Explicit treatment of the semicore electrons as valence electrons only reduces
$V_{xc}$ by 0.24 eV. The reduction in $V_{xc}$ is partially compensated
by the decrease of $\Sigma_{x}$ (0.08 eV) and $\Sigma_{c}$
(0.04 -- 0.15 eV). Consequently the overall change of $\Delta E_{VBM}$
is less than 0.1 eV. 
On the other hand, the conduction band minimum (CBM) orbital is
almost purely of Ti $3d$ character with substantial overlap to the semicore electrons.
The expectation value of ($V_{xc}$)
changes by about 9 eV.  Of course, there is a corresponding, large change in the
pseudopotential between these two cases.
The bare exchange term in the electron self energy operator changes by a similar amount.
Their combined
contribution to $\Delta E_{CBM}$ is only about -0.34 eV. 
The changes in the correlation part of the self energy $\Sigma_{c}$
of the CBM orbital is sensitive to the GW method. For the full-frequency
method, $\Sigma_{c}$ is decreased by only 0.1 eV when semicore electrons
are included in the PP, so we find that most of the net effect on the band gap
comes from the difference between LDA and bare exchange interactions
with the semicore electrons.

In Table \ref{tab:1} the results of using different methods 
to address the frequency dependence of W are shown. 
These affect the correlation part of the electron self energy
$\Sigma_{c}$ and the renormalization factor $Z$.
The results obtained with the full-frequency dependent 
dielectric function, evaluated using the contour deformation method
are the reference results, designated FF(CD) in the table.
For comparison, results from two different generalized plasmon pole models are shown.
In the Hybertsen-Louie approach\cite{Hybertsen1986}, designated GPP(HL), sum rules are applied
to each individual dielectric matrix element to develop a plasmon-pole model for its
frequency dependence.  
As shown in the lower part of Table \ref{tab:1}, when the semicore electrons of Ti were explicitly included in the pseudopotential, the calculated energy gap is 
almost 1 eV too large.
In the approach of von der Linden and Horsch\cite{Linden1988}, designated GPP(VDLH),
each dielectric matrix is first transformed to the basis of eigenpotentials 
and then sum rules are applied to develop a plasmon pole model for each eigenpotential.
This model also overestimates the energy gap, but by a smaller amount.
The renormalization factor $Z$, in addition to entering the evaluation of the quasiparticle energies in Eq. (\ref{eq:4}),
indicates the degree of dynamical admixture of collective excitations into the quasiparticle.
Larger deviations from unity indicate more admixture.
Compared with the FF results, the value of $Z$ indicates that the degree of admixture
is substantially underestimated by both GPP models. 

We have also tested the GPP(HL) approximation for
anatase TiO$_{2}$ as well as two other titanates,
SrTiO$_{3}$ and BaTiO$_{3}$.
In all three cases, the GPP(HL) approximation leads
to minimum band gaps that are too large by 0.7-0.8 eV.
A similar deviation for the renormalization constant, $Z$, is also observed. 
These results suggest that the quantitative issues with the plasmon pole
model extend more generally to titanates.
In previous calculations of the loss function \cite{Vast2002}
and the finite wavevector dynamical scattering factor \cite{Gurtubay2004}
for rutile TiO$_{2}$,
substantial structure is seen in the frequency dependence, well beyond
what could be easily accounted for by a single pole model.
These effects trace to an interplay between strong local field effects 
and the Ti semicore p- to empty d-shell excitation.
However, further analysis of the frequency dependence of the
screening in Si and LiCl shows that deviations from a pole
model for a range of frequencies around the plasmon energy is not
sufficient to predict the performance of the GPP model as it is used
to evaluate the GW expression.
The dynamical screening at larger frequencies only enters in an integrated fashion,
resulting in substantial cancellations internally.
In the case of titanates, we find that the full frequency dependence is essential
for quantitative results.
Similar conclusions were drawn for the case of metallic Cu \cite{marini2002b}.

\begin{table}
\begin{tabular}{cccccccccccccccccccc}
\hline 
\noalign{\vskip\doublerulesep}
 &  &  &  &  & \multicolumn{5}{c}{Rutile} &  &  &  &  &  &  & \multicolumn{3}{c}{Anatase} & \tabularnewline[\doublerulesep]
\noalign{\vskip\doublerulesep}
\noalign{\vskip\doublerulesep}
 & K-points &  &  &  & $\Gamma$ &  & $M$ &  & $R$ &  &  &  &  &  &  & $\Gamma$ &  & $X$ & \tabularnewline[\doublerulesep]
\hline
\noalign{\vskip\doublerulesep}
\noalign{\vskip\doublerulesep}
 & $E_{LDA}^{Val}$ &  &  &  & 0.00 &  & -1.06 &  & -1.04 &  &  &  &  &  &  & -0.48 &  & -0.05 & \tabularnewline[\doublerulesep]
\noalign{\vskip\doublerulesep}
\noalign{\vskip\doublerulesep}
 & $E_{GW}^{Val}$ &  &  &  & 0.00 &  & -1.15 &  & -1.12 &  &  &  &  &  &  & -0.58 &  & -0.06 & \tabularnewline[\doublerulesep]
\noalign{\vskip\doublerulesep}
\noalign{\vskip\doublerulesep}
 &  &  &  &  &  &  &  &  &  &  &  &  &  &  &  &  &  &  & \tabularnewline[\doublerulesep]
\noalign{\vskip\doublerulesep}
\noalign{\vskip\doublerulesep}
 & $E{}_{LDA}^{Cond}$ &  &  &  & 1.75 &  & 1.80 &  & 1.78 &  &  &  &  &  &  & 1.96 &  & 3.22 & \tabularnewline[\doublerulesep]
\noalign{\vskip\doublerulesep}
\noalign{\vskip\doublerulesep}
 & $E_{GW}^{Cond}$ &  &  &  & 3.38 &  & 3.40 &  & 3.34 &  &  &  &  &  &  & 3.56 &  & 4.89 & \tabularnewline[\doublerulesep]
\noalign{\vskip\doublerulesep}
\noalign{\vskip\doublerulesep}
 &  &  &  &  &  &  &  &  &  &  &  &  &  &  &  &  &  &  & \tabularnewline[\doublerulesep]
\noalign{\vskip\doublerulesep}
\noalign{\vskip\doublerulesep}
 & $E_{gap,GW}^{Direct}$ &  &  &  & 3.38 &  & 4.55 &  & 4.45 &  &  &  &  &  &  & 4.14 &  & 4.95 & \tabularnewline[\doublerulesep]
\noalign{\vskip\doublerulesep}
\noalign{\vskip\doublerulesep}
 & $E_{gap,GW}^{Indirect}$ &  &  &  & \multicolumn{5}{c}{3.34 ($\Gamma\rightarrow R$)} &  &  &  &  &  &  & \multicolumn{3}{c}{3.56 ($\Delta\rightarrow\Gamma$)} & \tabularnewline[\doublerulesep]
\noalign{\vskip\doublerulesep}
\noalign{\vskip\doublerulesep}
 &  &  &  &  &  &  &  &  &  &  &  &  &  &  &  &  &  &  & \tabularnewline[\doublerulesep]
\hline
\noalign{\vskip\doublerulesep}
\end{tabular}

\caption{LDA and quasiparticle energy levels of rutile and anatase near the
Fermi surface at selected k-points of high symmetry. Corresponding
quasiparticle energy gaps are also displayed. The energy reference
is taken to be the valence band maximum for both LDA and GW results. 
All energies are
in eV. \label{tab:2}}

\end{table}

The quasiparticle energy levels of rutile and anatase from the highest valence band
and the lowest conduction band
at selected high symmetry k-points are displayed in Table \ref{tab:2}. 
While the energy gap of rutile is found to be a direct gap (at $\Gamma$)
in the LDA, our FF GW calculations indicate it as indirect ($\Gamma\rightarrow R$).
However, the energy difference between the direct and indirect gap
is small. The energy of CBM at $\Gamma$ is only 0.04 eV higher than
the energy at $R$. 

\begin{figure}
\includegraphics[clip,scale=0.35]{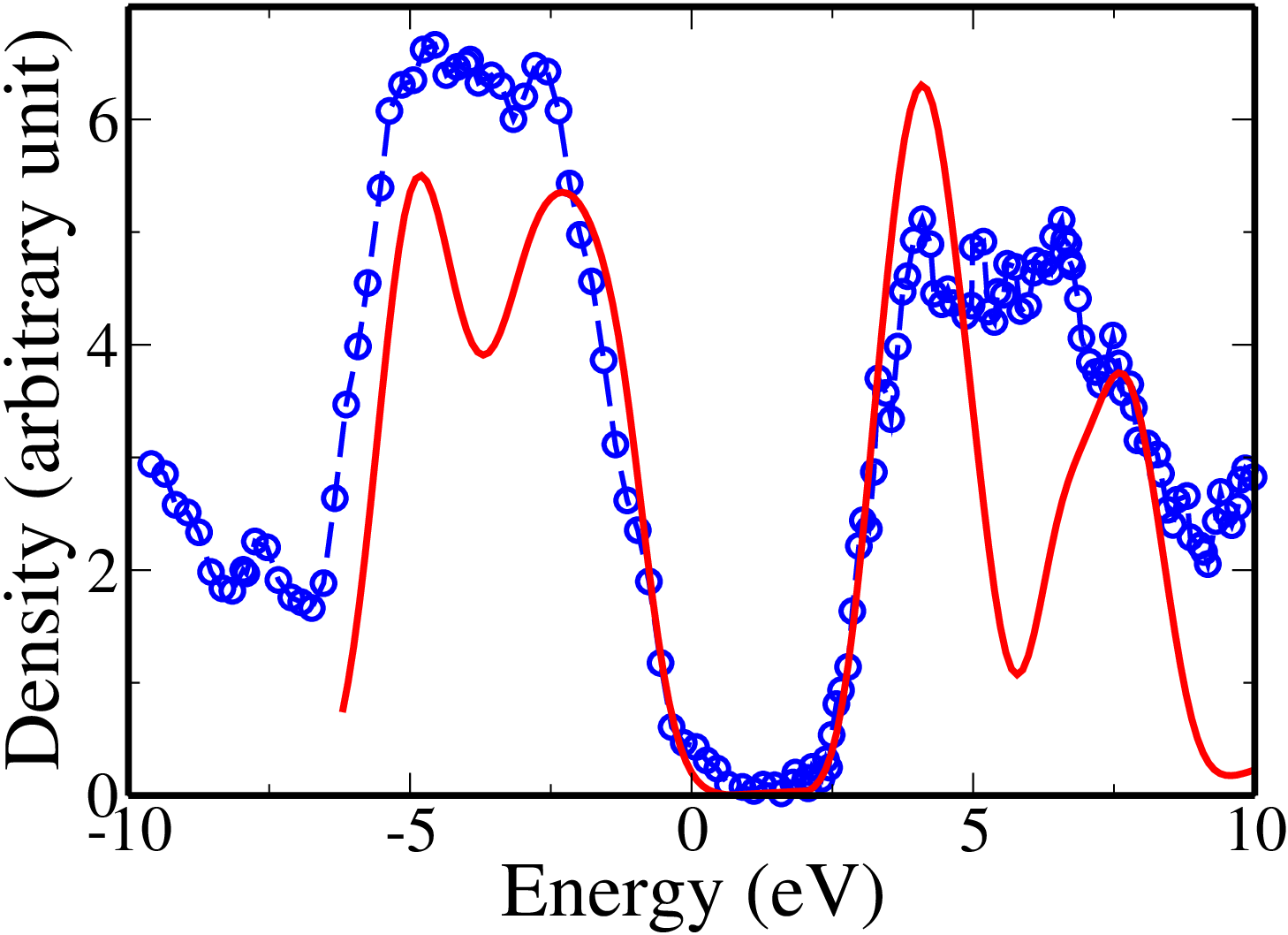}

\caption{Density of states (DOS) of rutile derived from FF GW calculations
plotted with photoemission and inverse photoemission spectra\cite{Tezuka1994}.
The solid curve is the calculated DOS which is convoluted with a Gaussian
function of $\sigma$= 0.5 eV, while circles are photo- and inverse
photo-emission spectra. \label{fig:1}}

\end{figure}

The calculated quasiparticle energies can be directly compared
to spectroscopic measurements for electron removal or addition to the solid.
The calculated value of the quasiparticle energy gap, 3.34
eV, agrees well with electron spectroscopy measurements, photo-emission and
inverse photo-emission measurements \cite{Tezuka1994, see1994}. 
In Fig. \ref{fig:1}, the density
of states (DOS) of rutile derived from FF GW calculations is plotted
together with the experimental spectra measured 
using x-ray photoemission and bremsstrahlung isochromat spectroscopy\cite{Tezuka1994}, which yielded a band gap of 3.3  $\pm$ 0.5 eV. 
Overall, the shape of the
calculated DOS matches the shape of the experimental spectra, especially
around the band gap.  
The experimental spectra measured using ultraviolet photoemission and inverse photoemission spectroscopy \cite{see1994} 
show very similar results with the inferred minimum energy gap about 0.2 eV smaller. 

The LDA calculations show that the top of the valence band of anatase
lies in the $\Delta$ direction, somewhere about 0.88 times of the
distance from $\Gamma$ to $X$. 
The energy at this k-point is 0.05
eV higher than the energy of the VBM at $X$. Subtracting the energy difference
as a perturbation from the quasiparticle energy gap between $X$ and
$\Gamma$, which is 3.62 eV from Table \ref{tab:2}, the quasiparticle
energy gap of anatase is found to be 3.56 eV.
The photoemission data available for anatase show an overall occupied
band with the oxygen p-bands \cite{Sanjines1994}.  That is similar to rutile.
To our knowledge, there is no inverse photoemission data available for anatase,
so the calculations can not be compared to a direct measurement of the quasiparticle energy gap.

\subsection{Optical Excitation Energies in TiO$_{2}$}

More precise measurements of the minimum energy gap
rely on optical absorption.
This introduces the extra complication that the
observed threshold for absorption will be altered
by interactions between the photoexcited electron and hole,
the formation of bound exciton states.
For rutile, the BSE calculation shows 
a series of singlet bound excitonic states at $\Gamma$. The lowest
two show $s$-state symmetry in the electron-hole envelope. They have a binding energy
of 0.13 eV and 0.06 eV respectively.  They are not dipole-allowed. The
third and forth are degenerate, with electron-hole envelope showing $p_{x,y}$ symmetry,
and have a binding energy of 26 meV. They are weakly dipole allowed for
the electric vector perpendicular to the c-axis. 
Together with the calculated direct quasiparticle gap energy from above,
we obtain the lowest energy singlet exciton at $\Gamma$ with energy 3.25 eV 
and the first dipole allowed singlet exciton with energy of 3.35 eV.
High resolution, low temperature optical absorption measurements
for rutile resolve several separate features \cite{Pascual1977}.
A very weak, but sharp exciton feature at 3.031 eV is identified as
the 1s exciton which is electric quadrapole allowed.  A stronger, 
but still relatively weak, dipole allowed 2p exciton
feature starts at 3.034 eV.  Finally, phonon-assisted features are also
identified that correspond to an indirect gap of 3.049 eV.

There are several important points of comparison.
First, the calculated lowest exciton energy at  $\Gamma$
is about 0.22 eV higher than measured.
Broadly, this error is comparable to those encountered when
using the GW approximation for other semiconductors \cite{Aryasetiawan1998,Aulbur19991,Onida2002}.
However, it is important to be clear that the calculation
is performed for a frozen lattice with no account given
for electron-phonon interactions.  In general, the electron-phonon
interaction will reduce the zero-temperature quasiparticle gap \cite{Cardona2005}.
Second, the present GW calculation gives the conduction band minimum
at  the R point instead of being at $\Gamma$, as suggested by the optical measurements.
The energy differences are small; the calculated conduction band at R is 0.04 eV lower than at $\Gamma$.
In the analysis of the absorption data, the indirect gap is found to be 0.01 eV higher 
than the direct gap \cite{Pascual1977}, albeit including what ever role excitonic effects may have.
The difference between theory and experiment is too subtle to be resolved in these calculations, particularly
without the influence of electron-phonon interactions.
Third, the symmetry of the excitonic states from the calculation agrees
with the interpretation of the experiments.
However, the scale of the excitonic effects that we calculate using the static
dielectric matrix, and only including the electronic polarizability,
is substantially larger than suggested by the experiment.
Our calculated long-wavelength dielectric constant ($\sim$ 8) is
slightly higher than the measured $\epsilon_{\infty}$ ($\sim$ 7) \cite{Cardona1965}, but similar to previous calculations \cite{Vast2002}.  
The lattice contribution is quite large, with $\epsilon_0$ = 111 \cite{Pascual1977}.
This again points to the importance of the electron-phonon interaction.

The measured optical absorption in single crystal anatase at low temperature
does not resolve any significant structure \cite{Tang1995}. 
The energy dependence near the onset of absorption is consistent
with an Urbach tail.  Analysis of the temperature dependence
leads to an estimate for the band gap for extended states of 3.42 eV \cite{Tang1995}.
This exceeds the measured optical threshold in rutile by about 0.4 eV.
Since the measured absorption edge in anatase is featureless,
another way to characterize the absorption edge and
make comparison to rutile is to consider the energy
at which the low temperature absorption is 50 $cm^{-1}$ 
for electric vector perpendicular to the c-axis.
In rutile, this occurs at 3.04 eV while it occurs at 3.30 eV in anatase  \cite{Tang1995}.
This suggests a more modest 0.26 eV difference between
the minimum energy gap of anatase and rutile.
Our calculated quasiparticle minimum energy gap in anatase is 3.56 eV, modestly 
larger than the value deduced from the absorption measurements.
The calculated difference in gaps between anatase and rutile is 0.22 eV,
similar to the measured difference.

So far, the calculated quasiparticle energies for TiO$_{2}$  have been found
to agree well with electron spectroscopy,
but the minimum energy gaps, including excitonic effects, are larger compared to 
the measured absorption threshold.
Furthermore, the strength of the calculated exciton binding energy
is larger than implied by the interpretation of the optical spectrum near threshold.
To get additional perspective, we calculate the
macroscopic dielectric function over a broad energy range,
including the correlations induced by electron-hole interactions.

\begin{figure}
\includegraphics[clip,scale=0.5]{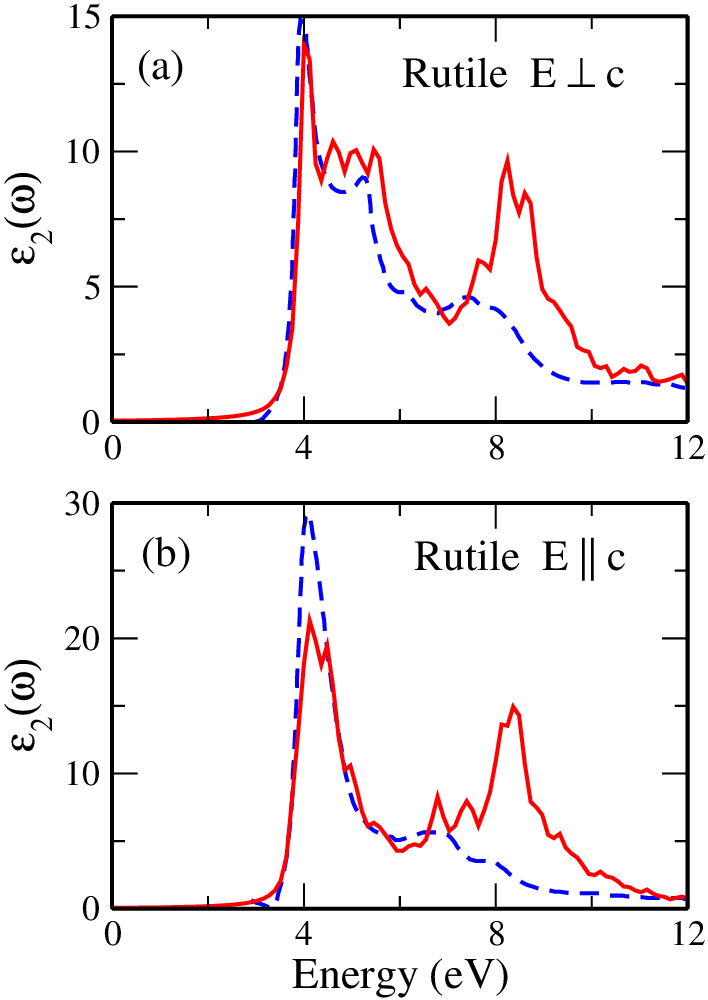}

\caption{$\epsilon_{2}(\omega)$ of rutile from 0 to 12 eV. Solid curves are
theoretical calculations with BSE, and dashed curves are experimental
results \cite{Cardona1965} obtained at room temperature. In (a) the
direction of polarization is perpendicular to the tetragonal axis
c, and in (b) the direction of polarization is parallel to axis c.\label{fig:2}}

\end{figure}

\begin{figure}
\includegraphics[clip,scale=0.5]{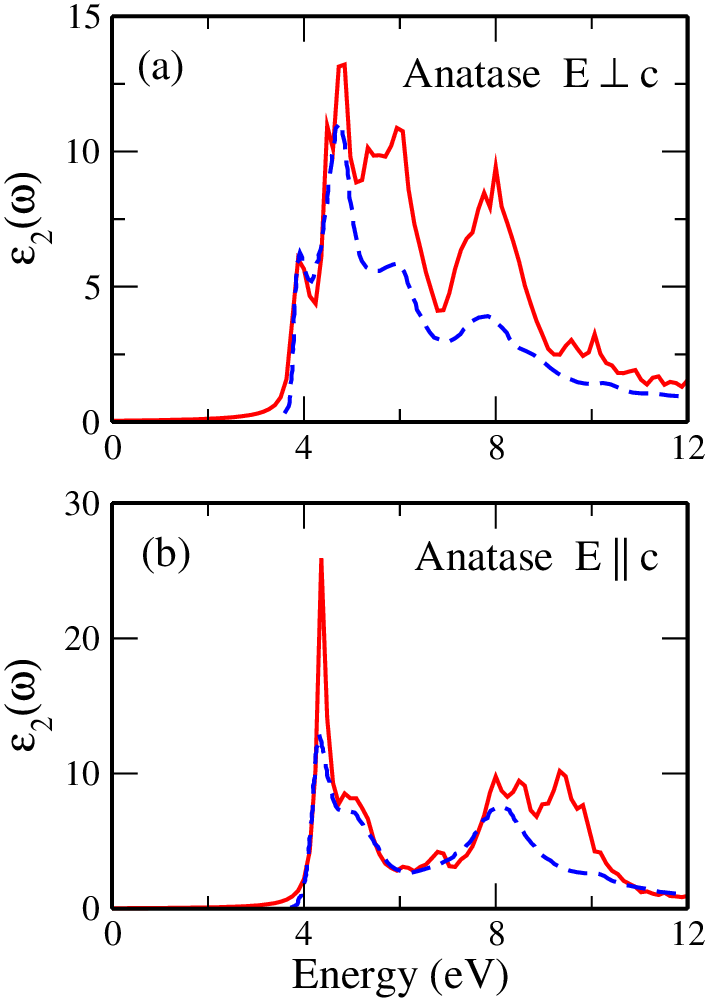}

\caption{$\epsilon_{2}(\omega)$ of anatase from 0 to 12 eV. Solid curves are
theoretical calculations from BSE, and dashed curves are experimental
results obtained at 100K \cite{Hosaka1997}. In (a) the direction
of polarization is perpendicular to the tetragonal axis c, and in
(b) the direction of polarization is parallel to axis c.\label{fig:3}}

\end{figure}

In Fig. \ref{fig:2}, we show the imaginary part of the dielectric
function of rutile for polarizations both perpendicular and parallel
to the tetragonal axis c. The solid curves are calculated from the BSE,
while the dashed curves are derived from optical reflectivity measurements
at room temperature \cite{Cardona1965}. For both polarizations, the
theoretical spectra are close to experiment up to about 6 eV. 
In particular, the first strong peak at $\sim$ 4 eV for both
polarizations is reproduced very well by the BSE results. 
Above 6 eV, the overall magnitude and prominence of the peaks
in theoretical spectra are distinct from experiment.
The $\epsilon_{2}(\omega)$ of anatase is displayed in Fig. \ref{fig:3},
where the experimental data \cite{Hosaka1997} were measured at 100
K. Similar to the rutile case, the theoretical calculations capture
the features around the onset of major absorption at 4 eV.
For perpendicular polarization, the calculated oscillator strength is systematically 
too large starting at about 5 eV.
The calculated results are very similar 
to the previous calculation for rutile TiO$_{2}$  \cite{Lawler2008}.
For anatase, the calculated peak heights near 4 eV appear less intense in their spectra,
but this is largely due to their choice of a larger damping parameter, as is evident from
the broadening on the low energy side of their spectra.
In particular, we have analyzed the integrated oscillator strength (i.e. the contribution
to the f-sum rule) from the first peak in the anatase spectrum for parallel polarization.  
We find that our oscillator
strength is essentially the same as theirs, but that both calculations show more
oscillator strength than is found in the experimental spectra by about 30\%.

The systematic overestimation of oscillator strength at higher energy appears
to be a more general issue.  For example, a recent BSE study for several alkaline earth metal monoxides shows some
similar excess oscillator strength at higher photon energies \cite{Schleife2009}.
This may well trace to more fundamental assumptions in the methodology.
Two key issues are the use of the Tamm-Dancoff approximation, which has been
identified recently as the main source of errors in the calculations
of a confined system \cite{Gruning2009}, and the assumption
of a statically screened Coulomb interaction \cite{Rohlfing2000, Marini2003}.
Also, as noted by Lawler and coworkers, the f-sum rule for the oscillator strength
converges very slowly in the titanates and the experimental analysis that relies
on Kramers-Kronig transformations may have systematic errors as well \cite{Lawler2008}.

\subsection{External Energy Level Alignment: TiO$_{2}$(110) Surface}

The energetic position of the quasiparticle VBM with respect to the vacuum level
in the rutile phase for the (110) surface was calculated in several steps. 
First at the LDA level, analogous to the determination of
work function for metals,
the electrostatic potential change between the bulk like region and the vacuum region
was determined  \cite{Zhang1988,Baldereschi1988,Fall1999,Shaltaf2008}. 
Then in a second step, the bulk VBM position relative to the electrostatic potential
is determined.
Previous calculations of work functions based on metal slabs
based on the DFT Fermi energy 
were in fairly good agreements with experimental measurements \cite{Fall1998,Fall2000}. 
In the third step, we will apply the perturbative correction to the LDA
exchange-correlation potential to determine the alignment 
in the GW approximation  \cite{Zhang1988,Shaltaf2008}.

For semiconductors and insulators the Fermi level is separated from
the VBM, and the position of VBM varies as the surface structure changes.  
To calculate the energetic alignment of rutile with respect to the
vacuum from the (110) surface, a stoichiometric slab of six layers
of Ti atoms is cut from the bulk structure with geometrical parameters
described in Sec. I. A vacuum space of the same thickness of the slab
is used to buffer the two surfaces of the slab. The surface is cleaved
as a (1$\times$1), unreconstructed surface as observed in experiments \cite{Diebold2003}.
The k-point mesh is $2\times4\times1$. After the slab is fully relaxed
with the atomic positions of the central two layers fixed, the electrostatic
potential inside the slab with respect to the vacuum is measured as
-12.65 eV. 
The energy difference between the VBM (from the LDA calculation) and
the electrostatic potential inside the material is 5.05 eV, which
is determined in a separate bulk calculation to prevent the quantum
size effects \cite{Fall1998}.  Accordingly, the LDA value for the VBM
relative to vacuum is 7.60 eV.  Previous LDA and gradient corrected
calculations for  give 7.16 eV (LDA, three Ti layer slab) \cite{Vogtenhuber2002}, 
7.2 eV (PBE, 11 Ti layer slab)  \cite{Kiejna2006} and 7.6 eV (PW91, 11 Ti layer slab) \cite{Kiejna2006}.
In light of variations at the 0.2 eV level with number of layers in the slab model \cite{Kiejna2006},
the overall agreement is satisfactory.

The GW correction for the VBM calculated with FF model and 160 bands
is 0.07 eV. However, as noted in Sect. IIC, extrapolation to full
convergence with respect to the total number of bands will drive this
 0.2 to 0.4 eV lower. We therefore
suggest a GW correction of  -0.2 eV,
with about 0.1 eV uncertainty. The final prediction from GW for the
VBM alignment to vacuum at the clean rutile TiO$_{2}$ (110) surface
is  7.8 eV.

In order to deduce the VBM alignment form experiment, two
results must be combined: (1) the work function which fixes the Fermi energy
relative to the vacuum; (2) the position of the VBM relative to the Fermi energy.
Experimental measurements of the work function of rutile from the
(110) surface \cite{Onda2004,Schierbaum1996} vary from 4.7 eV to
5.8 eV depending on the structure of the surface which is strongly influenced
by treatment (annealing, exposure to oxygen, etc.). 
In addition, the position of the VBM relative to the Fermi energy also
depends on surface treatment \cite{Wendt2008}.
Therefore, it is crucial to compare with data in which both values are
measured on the same sample.
To our knowledge, this is relatively rare.
Based on UPS measurements with ($h \nu$=21.2 eV), a work function of 5.2 eV
and a relative VBM position of 3 eV were measured for TiO$_{2}$(110) \cite{Schierbaum1996},
which implies the VBM position relative to vacuum falls at 8.2 eV.
Similar measurements for TiO$_{2}$(100) yield a work function of 4.9 eV and a relative VBM position of 3.1 eV respectively \cite{Xiong2007},
yielding the VBM position at 8.0 eV.
The difference in workfunction between these measurements is consistent with 
a separate Auger Microprobe study of facet dependence \cite{Imanishi2007}.

Based on this limited data set, the GW based prediction for the VBM
alignment is off by about 0.4 eV.
There are at least 0.1 eV uncertainties in both the theory and the
experiment. Since both the work function and position of Fermi level
are sensitive to the surface properties, the deviation between the
theoretical calculation and experimental measurements may well reflect
the complexity of the TiO$_{2}$ surfaces. For example, recent studies
suggest that the commonly employed strategy of cleaning followed by
annealing in oxygen may not result in the ideal surface envisioned
(e.g. with no oxygen related defects) \cite{Wendt2008}.
In particular, the physical origin of the widely observed defect states
around 1 eV below the CBM remains a point of vigorous discussion\cite{DiValentin2006,
Wendt2008}.

\section{Concluding Remarks}

We have presented a numerically well converged MBPT study of
the electronic and optical excitation energies in rutile and anatase
crystals of TiO$_{2}$.  The calculations are carried out in
the approximation of a frozen lattice, without consideration of electron-phonon coupling.  
In most respects, the agreement with
experiment is well within the expected accuracy of this approach.  
In particular, the calculated quasiparticle gap agrees with electron
spectroscopy measurements (photoemission and inverse photoemission),
the change in the minimum gap between rutile and anatase crystal
structures is reproduced, 
and the main features of the optical spectrum agree with ellipsometry measurements.
The qualitative features of the zone center bound excitons calculated
for rutile agree well with low temperature absorption measurements.
However, the scale of the exciton binding energy is larger than that estimated from
experiment by about a factor of 10 and the calculated exciton energy is about
0.2 eV larger than measured.

The key theoretical assumptions in our application of MBPT include
evaluation of the electron self energy in the GW approximation without
iterating to self consistency or considering vertex corrections.
Self consistency would certainly increase the calculated energy gap
through the reduction in the screening
\cite{Schilfgaarde2006,Shishkin2007b}.
Recent results for a set of other semiconductors and insulators shows that
approximate inclusion of vertex corrections in screening leads to
a partially compensating reduction of the calculated energy gap\cite{Shishkin2007a}.
However, a fully
consistent approach
to include vertex corrections remains subject of current research in the field.
Because the Ti $3d$ electrons are almost completely ionized,
TiO$_2$ should not be subject to the sorts of systematic errors
found in non-selfconsistent calculations for some late transition $3d$ metal 
compounds\cite{Aryasetiawan1995,Faleev2004,Bruneval2006a}
The accurate results found for the key optical transition energies
contributing to the absorption (Figs. 3 and 4) support this,
and contrast to the case of Cu$_2$O where non-selfconsistent
calculations showed substantial discrepencies\cite{Bruneval2006a}.
In the solution of the BSE, the calculated static (electronic) dielectric
matrix has been used and the equations were simplified through the Tamm-Dancoff approximation.
These approximations are part of the standard
MBPT treatment of optical spectra and
the low energy excitons are expected to be treated well\cite{Rohlfing2000,Onida2002}.
However, they may affect higher energy features in the spectra\cite{Marini2003}.

On physical grounds, we suggest that the most significant open issue
concerns the role of electron-phonon coupling.
In general, the electron-phonon self energy
will both lead to a smaller zero-temperature quasiparticle gap
and make a significant contribution to the temperature dependence of the energy gap \cite{Cardona2005}.
As noted in the text, there is a very large difference between 
the electronic dielectric constant $\epsilon_{\infty}$ and
the low frequency dielectric constant including lattice polarization $\epsilon_0$
for TiO$_{2}$. 
This suggests a relatively strong electron-phonon interaction
and there is a long standing debate over the polaronic character
of charge excitations in TiO$_{2}$ \cite{PNote}.

Based on the usual form of the Frohlich interaction, the dimensionless
coupling constant characterizing electronic coupling to the most important
polar optic mode for rutile (with mode energy about 0.1 eV)
is given by $\alpha=1.6\sqrt{m_{b}/m_{e}}$ where $m_{b}$ is the
bare band mass and $m_{e}$ is the free electron mass.
Using our DFT band dispersions to have estimates, the electron (hole)
band mass is about 0.6$m_{e}$ (1.8$m_{e}$) along $x$ or $y$ and
1.6$m_{e}$ (3.1$m_{e}$) along $z$. This suggests coupling constants
of $\alpha\sim1-2$ for electrons and $\alpha\sim2-3$ for holes which
would fall in the weak to intermediate coupling regime. Using the
usual weak coupling expression, the electron and hole renormalization
would be $0.1-0.2$ eV and $0.2-0.3$ eV respectively, both of which
act to reduce the quasiparticle energy gap. For anatase $\alpha\sim1.6\sqrt{m_{b}/m_{e}}$,
essentially the same as rutile, based on the mode energy and dielectric constants
\cite{Gonzalez1997}. The electron
(hole) band mass is about 0.4$m_{e}$ (1.8$m_{e}$) along $x$ or
$y$ and 3.9$m_{e}$(1.0$m_{e}$) along $z$, suggesting slightly different coupling
constants of $\alpha\sim1-3$ for electrons and $\alpha\sim2$ for
holes with corresponding (weak coupling) electron and hole renormalization
of 0.1-0.3 eV and 0.2 eV respectively. 
In weak coupling, the electron-phonon self energy is added to the results obtained
here based on the GW approximation.
For the analysis of the optical absorption edge, a more detailed calculation
is required because in the exciton-phonon coupling, the exciton is neutral and
the electron and hole distortions of the lattice
will partially cancel \cite{SchmittRink1987, Rudin1990}.
Taken together, if the large polaron regime is physically correct,
these rough estimates suggest that the effect of the electron-phonon coupling could
account for some of the differences between the present GW/BSE
results for the frozen lattice and experiment. 
Firmer conclusions require a more extensive set of calculations,
beyond the scope of this article
\cite{Cardona2005,Marini2008,Bechstedt2005,Vidal2010}.
In particular, it may be that a more complete consideration of self-consistency
and vertex corrections in the electron-electron contribution to the electron self energy will need
to be combined with an analysis of the electron-phonon contribution.
The two contributions should be treated
in a fully consistent theory.
In more empirical terms, an overstimate of the band gap  based on
selfconsistent treatment of the electron-electron
interactions alone may be compensated by the electron-phonon contributions.

It  may well be the case that the large polaron regime is not applicable for TiO$_{2}$.
A recent THz spectroscopy study
of rutile gave a direct measurement of the electron scattering rate  \cite{Hendry2004}.
This data was analyzed with a Frohlich form for the electron-phonon interaction,
but regarding the coupling constant as a free parameter. 
Using the Feynman approach \cite{Feynman1955,schultz1959} to handle intermediate
to strong coupling, the analysis showed coupling constants for electrons 
$\alpha \sim 4 - 6$ depending on field orientation \cite{Hendry2004}.  
The inferred electron mobilities were consistent with earlier electron transport measurements \cite{Yagi1996}.
These values suggest a substantially larger value for the
electron self energy of $0.4 - 0.7$ eV \cite{schultz1959}.  An older estimate based on
small polaron theory also suggested 0.7 eV \cite{Eagles1964}.  A recent DFT+U based study
suggested that an excess electron in rutile is indeed self trapped \cite{deskins2007}.  Although 
the binding energy was not presented, the barrier for polaron hopping
was estimated to be 0.3 eV. 
A similar study for an excess hole in rutile
suggested barriers of 0.5 - 0.6 eV \cite{Deskins2009}.
Taken together, if the small polaron regime is found to be physically relevant,
then the quasiparticle and excitonic energies will need to be fully reanalyzed.
For strong electron-phonon coupling a perturbative 
approach to combine the electron-electron and electron-phonon self energies is no longer justified.
Furthermore, the electron-phonon coupling enters into the spectroscopic measurements
in distinct ways.  The (inverse) photoemission and optical absorption would each need to be
properly analyzed.

\begin{acknowledgments}
We thank Dr. A. Marini for access to the private
branch of the Yambo code.
W. K. thanks Dr. D. Prezzi for insightfull discussions on BSE.
Work performed under the auspices
of the U.S. Department of Energy under Contract No. DEAC02-
98CH1-886. This research utilized resources 
at the New York Center for Computational Sciences 
at Stony Brook University/Brookhaven National Laboratory 
which is supported by the U.S. Department of Energy under 
Contract No. DE-AC02-98CH10886 and by the State of New York.
\end{acknowledgments}
\bibliography{titanium_article_bib_database}

\end{document}